\documentclass{article}
\usepackage{microtype}
\usepackage{graphicx}
\usepackage{subfigure}
\usepackage{booktabs} % for professional tables

\usepackage{hyperref}
\usepackage{wrapfig}
\usepackage[T1]{fontenc}

\usepackage{aas_macros} 

\usepackage[accepted]{icml2024}

\usepackage{amsmath}
\usepackage{amssymb}
\usepackage{mathtools}
\usepackage{amsthm}

\usepackage[capitalize,noabbrev]{cleveref}

\usepackage{algorithm}
\usepackage{algorithmic}

\usepackage[textsize=tiny]{todonotes}

\usepackage{typed-checklist}

\theoremstyle{plain}

\theoremstyle{definition}

\theoremstyle{remark}

\icmltitlerunning{Dark Energy Constraints using Simulation-Based Inference}
\begin{document}

\twocolumn[
\icmltitle{Population-level Dark Energy Constraints from Strong Gravitational Lensing using Simulation-Based Inference}

\icmlsetsymbol{equal}{*}

\begin{icmlauthorlist}
\icmlauthor{Sreevani Jarugula}{fermilab}
\icmlauthor{Brian Nord}{fermilab,uchi}
\icmlauthor{Abhijith Gandrakota}{fermilab}
\icmlauthor{Aleksandra Ćiprijanović}{fermilab,uchi}
\end{icmlauthorlist}

\icmlaffiliation{fermilab}{Fermi National Accelerator Laboratory}
\icmlaffiliation{uchi}{Kavli Institute for Cosmological Physics \& Department of Astronomy and Astrophysics, The University of Chicago}

\icmlcorrespondingauthor{Sreevani Jarugula}{jarugula@fnal.gov}

\icmlkeywords{Machine Learning, ICML}

\vskip 0.3in
]

\printAffiliationsAndNotice{\icmlEqualContribution}

\begin{abstract}

In this work, we present a scalable approach for inferring the dark energy equation-of-state parameter ($w$) from a population of strong gravitational lens images using Simulation-Based Inference (SBI). 
Strong gravitational lensing offers crucial insights into cosmology, but traditional Monte Carlo methods for cosmological inference are computationally prohibitive and inadequate for processing the thousands of lenses anticipated from future cosmic surveys. 
New tools for inference, such as SBI using Neural Ratio Estimation (NRE), address this challenge effectively. 
By training a machine learning model on simulated data of strong lenses, we can learn the likelihood-to-evidence ratio for robust inference. 
Our scalable approach enables more constrained population-level inference of $w$ compared to individual lens analysis, constraining $w$ to within $1\sigma$. 
Our model can be used to provide cosmological constraints from forthcoming strong lens surveys, such as the 4MOST Strong Lensing Spectroscopic Legacy Survey (4SLSLS), which is expected to observe 10,000 strong lenses. 
\end{abstract}

%%%%%%%%%%%%%%%%%%%%%%%%%%%%%%%%%%%%%%%%%%%%%%%%%%%%%%%%%%%%%%%%%%%%%
%%%%%%%%%%%%%%%%%%%%%%%%%%%%%%%%%%%%%%%%%%%%%%%%%%%%%%%%%%%%%%%%%%%%%
\section{Introduction}
\label{sec:introduction}

Dark energy, which comprises approximately 70$\%$ of the energy density of the universe, plays a pivotal role in driving the accelerated expansion of the universe. 
Yet, the nature of dark energy remains a prominent puzzle in physics. 
Central to understanding its fundamental nature is the dark energy equation-of-state parameter $w = p/ \rho$, the ratio between the dark energy pressure $p$ and the energy density $\rho$. 
High-precision constraints on $w$ are crucial for understanding the fate of the expansion of the universe \citep[e.g.,][]{caldwell03, sahni02}. Constraints on dark energy have been provided through observations of the Cosmic Microwave Background~\citep[CMB;][]{planck20}, supernova \cite{riess16}, and Baryon Acoustic Oscillations~\citep[BAO;][]{alam17} which are in broad agreement with the constant value of $w=-1$ according to the $\Lambda$ cold dark matter ($\Lambda$CDM) cosmological model \cite{escamilla23}. 
However, $w$ is degenerate with other cosmology parameters such as the Hubble constant $H_{0}$ and the dark matter density $\Omega_{m}$, necessitating careful statistical analysis combining multiple observational probes. Given the existing tensions in cosmology parameter estimates using probes from the early and late universe observations \citep[e.g.,][]{leizerovich23}, improving constraints on $w$ using independent cosmological probes like strong gravitational lensing is crucial.

The dark energy equation-of-state parameter can be constrained through the distance ratio from static galaxy-galaxy strong lens systems by combining lensing and stellar dynamical measurements. 
The constraints on $w$ have been obtained from a population of $\mathcal{O}(100)$ strong lens systems \citep{cao15, jie16} using Markov Chain Monte Carlo (MCMC) with analytic likelihoods \cite{lewis02}. 
However, traditional MCMC methods are limited by the need for accurate and efficient lens modeling with numerous parameters, making it computationally prohibitive to analyze large datasets. 
More efficient techniques to model strong lensing systems are required in the upcoming era of Legacy Survey of Space and Time~\citep[LSST;][]{ivezic08}, Euclid Wide Survey \cite{euclid22}, and Roman Space Telescope High Latitude Wide Area Survey \cite{spergel15} where $\mathcal{O}(10^{5})$ strong lens systems will be discovered \cite{holloway23}. 
Spectroscopic information from surveys such as the 4MOST Strong Lensing Spectroscopic Legacy Survey~\citep[4SLSLS;][]{collett23}, expected to observe about 10,000 strong lenses, will provide better constraints on $w$ \cite{litian24}.
However, the analyses so far are limited by the complexity of lens modeling in calculating the analytical likelihood, highlighting the need for faster and scalable inference algorithms.

AI-based methods such as Simulation-Based Inference (SBI) using Bayesian Neural Networks (BNN), Neural Posterior Estimation (NPE), and Neural Ratio Estimation (NRE) have emerged as powerful methods for posterior inference without explicitly calculating the likelihood \cite{cranmer20}. 
These methods have been successfully applied in various astrophysical and cosmology studies \citep{brehmer19, legin21, legin22, gerardi21, wagner21, zhang22, khullar22, poh22, mishra22, wagner23, moser24,lemos24}. 
The NRE method involves training a neural network on simulations of data to estimate the
likelihood-to-evidence ratio \cite{hermans20} which can then be used to perform approximate Bayesian inference. This approach allows for efficient posterior inference without the need for explicit likelihood evaluations, making it particularly useful for complex simulations with intractable likelihoods. Moreover, NRE facilitates population-level inference, allowing for inference on parameters common across a population while marginalizing over variables included in the simulation but not of inferential interest (nuisance parameters). A key advantage of these SBI methods is their ability to amortize the computational cost of the inference procedure where after an upfront cost of simulation and neural network model training, efficient inference can be performed on a large number of observations. 

This analysis presents the first application of NRE for population-level inference, constraining the dark energy equation-of-state parameter $w$ from strong gravitational lens images. We estimate the posterior distribution of $w$ from the likelihood-to-evidence ratio from NRE using MCMC sampling and analytical calculations. After introducing strong gravitational lensing and the simulation setup (Section \ref{sec:background}), we delve into the background of NRE and details of posterior calculation methods (Section \ref{sec:sbi}). Details on the training and test data used for the NRE model are then provided in Section \ref{sec:data}. 
Section \ref{sec:results} focuses on the analysis and results. We evaluate the performance of the trained NRE model by comparing its predicted values of $w$ to the true values. Furthermore, we present the population-level inference results, showcasing the ability of NRE to constrain $w$ simultaneously from multiple lens systems. Finally, in Section \ref{sec:conclusion}, the key findings are summarized, accompanied by an outlook on potential future analyses.

%%%%%%%%%%%%%%%%%%%%%%%%%%%%%%%%%%%%%%%%%%%%%%%%%%%%%%%%%%%%%%%%%%%%%
%%%%%%%%%%%%%%%%%%%%%%%%%%%%%%%%%%%%%%%%%%%%%%%%%%%%%%%%%%%%%%%%%%%%%
\section{Strong Gravitational Lensing}
\label{sec:background}
Strong gravitational lensing is a phenomenon in which the mass distribution of a galaxy (the lens) distorts the light from a background source, producing distorted and magnified images of the source in the observed image plane. 
This phenomenon arises due to the multiple solutions to the lens equation. 
In the case of a point-like background source, such as a quasi-stellar object (QSO) or a supernova, the lensing effect can lead to the formation of multiple images of the source. 
In the case of an extended background galaxy, lensing results in an arc-like structure in the image plane, which we refer to as galaxy-galaxy lensing. 
Future surveys are expected to discover thousands of galaxy-galaxy strong lenses making this increasingly relevant for high-redshift and cosmology studies. 
Notably, the strong lensing observations can constrain the dark energy equation-of-state parameter $w$, which plays a crucial role in our understanding of the expansion of the universe. 
The constraint on $w$ can be derived from the analysis of the distance ratio, defined as the ratio of the angular diameter distance between the lens and the source, and the distance between the source and the observer. 
In this section, we present some of the equations that are used in the simulator.

The lens equation relates the source position $\beta$ to the lensed source position $\theta$ through the deflection angle $\alpha(\theta)$ as 
\begin{equation}
    \beta\ = \theta\ - \alpha(\theta).
\end{equation}
% The deflection field is the gradient of the lensing potential ($\psi$) 
% \begin{equation}
%     \alpha(\theta)\ = \nabla \psi(\theta)\ = \frac{1}{\pi}\int d\theta^{\prime}\ \frac{\theta\ - \theta^{\prime}}{|\theta\ - \theta^{\prime}|^{2}}\kappa(\theta^{\prime}),
% \end{equation}
The deflection field is described by the projected surface mass density $\kappa(\theta)\ = \Sigma(\theta)/\Sigma_{cr}$.
The critical lensing surface density $\Sigma_{cr}$ is given by 
\begin{equation}
    \Sigma_{cr} = \frac{1}{4\pi\ G}\frac{D_{s}}{D_{ls}D_{l}}, 
\end{equation}
where $D_{l}$, $D_{s}$, and $D_{ls}$ are the angular diameter distances between the observer and the lens, the observer and the source, and the lens and the source, respectively. 
In $w$CDM cosmology, the angular diameter distance depends on the astrophysical (local) parameters of redshift for the source and the lens and on cosmology (population-level) parameters --- Hubble constant $H_{0}$, matter density parameter $\Omega_{m}$, dark energy density parameter $\Omega_{de} = 1 - \Omega_{m}$, and the dark energy equation-of-state parameter $w$:

% \begin{equation} \label{distance_equation}
% \begin{split}
%         D(z, H_{0}, \Omega_{m}, w) = \frac{1}{1+z}\frac{c}{H_{0}}\int_{0}^{z}\frac{dz^{\prime}}{h(z^{\prime},\Omega_{m}, w)},\\
%         h^{2}(z,\Omega_{m}, w) = \Omega_{m}(1+z)^{3} + (1-\Omega_{m})(1+z)^{3(1+w)}.
% \end{split}
% \end{equation}

\begin{equation} 
\label{distance_equation}
        D(z, H_{0}, \Omega_{m}, w) = \frac{1}{1+z}\frac{c}{H_{0}}\int_{0}^{z}\frac{dz^{\prime}}{h(z^{\prime},\Omega_{m}, w)},
\end{equation}
where the Hubble parameter is
\begin{equation} 
        h^{2}(z,\Omega_{m}, w) = \Omega_{m}(1+z)^{3} + (1-\Omega_{m})(1+z)^{3(1+w)}.
\end{equation}

% The gravitational lensing formalism is described in detail in \citet{treu10}.

% The Einstein radius for the Singular Isothermal Spheroid (SIS) profile, which is a specific case of the SIE profile is described in \citet{treu10} as
The Einstein radius is given by \citep{treu10}: 
\begin{equation}
    \theta_{E}\ = 4\pi \left(\frac{\sigma_{v}}{c} \right)^{2} \frac{D_{ls}}{D_{s}}.
\end{equation}
Through observations of the Einstein ring from strong lensing and stellar velocity dispersion from dynamical studies, the cosmology ($\Omega_{m}, w$) can be constrained through the distance ratio $\frac{D_{ls}}{D_{s}}$.

We model the mass of the lens using a Singular Isothermal Ellipsoid (SIE) profile \cite{kormann94}. 
The convergence of the lens is described by  $\theta_{E}$, lens position ($x_{l}$, $y_{l}$), and the ellipticity of the lens ($l_{e1}, l_{e2}$). 
We model the light profile of the source using a Sersic profile \citep{sersic68} which is described by the magnitude of the source ($m_{s}$) at the effective radius $R$, the source position ($x_{s}$, $y_{s}$), Sersic index $n$, and the source ellipticity ($s_{e1}, s_{e2}$).

\section{Simulation-based Inference}
\label{sec:sbi}

Given a population of strong lens images, our goal is to infer the dark energy equation-of-state parameter $w$. 
In the context of population-level inference, we implement an approach akin to hierarchical Bayesian modeling, where we combine information from a sample of observations to obtain tighter constraints on the parameters that are common across the population.
% The simulation setup as described in Section \ref{sec:simulation_setup} generates images through forward modeling given the parameter of interest $w$ and the nuisance variables $z$ where the images $x \sim\ p(x|w)$. 
In Bayesian inference, the posterior is given by 
\begin{equation} \label{eqn:bayes}
    p(w | x) = \frac{p(x|w)}{p(x)}p(w),
\end{equation}
where $p(w)$ is the prior on $w$, $p(x|w)$ is the likelihood of the data $x$ given $w$, and $p(x)$ is the evidence.
The likelihood depends on nuisance parameters ($\nu$) that we integrate over:
\begin{equation}
    p(x | w) = \int p(x, \nu|w) d\nu,
\end{equation}
where $p(x, \nu|w)$ is the combined likelihood of observed variables and nuisance parameters. 
In many scenarios in modern cosmology, the likelihood is intractable because of the large number of nuisance parameters, which require a high-dimensional integral. 
% and the integration can not be computed explicitly. 
% Simulations from a forward model simulator as described in Section \ref{sec:simulation_setup} can be used to learn the likelihood function or the posterior directly using a neural network \cite{cranmer20}. 
In this work, we employ NRE to obtain the posterior $p(w | x)$ \citep{cranmer15, baldi16}. 
With NRE, a classification neural network is trained to model the likelihood ratio between two hypotheses. 
Each hypothesis represents a probability distribution from which the samples ($x, w$) are drawn. 
If one of the hypotheses is that the samples are drawn from $p(x|w)$ and the other is $p_{\rm ref}(x|w)$, the likelihood ratio $r(x|w) = \frac{p(x|w)}{p_{\rm ref}(x|w)}$. 
Using a Binary Cross Entropy Loss (BCE) function, the optimal classifier that differentiates the samples drawn from these two hypotheses is given by 
\begin{equation}
    d^{*}(x, w) = \frac{p(x|w)}{p(x|w) + p_{\rm ref}(x|w)}.
\end{equation}
The ratio $r(x|w)$ is related to the classifier $d^{*}(x, w)$ as
\begin{equation}
    r(x|w) = \frac{d^{*}(x, w)}{1 - d^{*}(x, w)}.
\end{equation}
The classification neural network learns to map the $(x, w)$ samples to the class probabilities through its final layer. 
The layer preceding the final activation outputs the \textit{logit}, which is the logarithm of the likelihood ratio log\,$r(x|w)$.

In our analysis, we use a classifier to distinguish between the sample-parameter pairs: $(x, w) \sim\ p(x, w)$ drawn from the joint distribution with class label $y=1$ and $(x, w) \sim\ p(x)p( w)$ drawn from the marginal distribution with class label $y=0$. 
If we consider the two hypotheses to be the joint and the marginal distribution, the network learns the likelihood-to-evidence ratio $r(x|w) = \frac{p(x, w)}{p(x)p( w)}= \frac{p(x|w)}{p(x)}$ \cite{hermans20}.

The joint likelihood-to-evidence ratio from a population of lenses $\{x\}$ can be obtained by combining the likelihood ratio from individual lenses under the assumption that the observations are independent and identically distributed. 
This joint likelihood ratio can be written as

\begin{equation} \label{eqn:joint_likelihood}
r(\{x\} | w) = \prod_i\ r(x_i | w) = \sum_{i} log\  r(x_i | w).
\end{equation}

An important advantage of estimating the likelihood-to-evidence ratio is its ability to combine the information from a population of observations that have the same underlying parameter of interest. 
This provides more robust constraints on the population-level parameters \citep{brehmer19, zhang22}.

% Using a Binary Cross Entropy Loss (BCE) function, the optimal classifier is 
% \begin{equation}
%     d^{*}(x, \theta) = \frac{p(x,\theta)}{p(x,\theta) + p(x)p( \theta)}.
% \end{equation}
% The ratio $r(x|\theta)$ is related to the classifier $d^{*}(x, \theta)$ as
% \begin{equation}
%     r(x|\theta) = \frac{d^{*}(x, \theta)}{1 - d^{*}(x, \theta)}
% \end{equation}
% The classifier is trained through supervised learning and the log\,$r(x|\theta)$ is obtained by reading the \textit{logits} (pre-sigmoid) outputs of the model.

\subsection{Neural Ratio Estimation Model Architecture}
To learn the likelihood ratios, we use a ResNet architecture \cite{he15}.
The model architecture is shown in Table \ref{tab:network_summary_skip} in Appendix \ref{sec:app:network}. 

We implement the model in \textit{TensorFlow} \cite{tensorflow2015-whitepaper}.
The inputs to the model are the images and the corresponding parameter of interest ($w$) that generated the images. 
As a pre-processing step, we normalize the images in the training data. 
The samples from the marginal distribution are generated in each batch, where $w$ is randomly selected for each image. 
Performing the randomization at the batch level gives the network more random samples to differentiate between the two distributions. 
Batch normalization is applied to the data followed by a convolutional layer with batch normalization and \textit{ReLU} activation function, serving as the entry block. 
Residual connections are introduced to preserve essential information across subsequent convolutional blocks of increasing depth and complexity. 
Each convolutional block consists of two convolutional layers with batch normalization and \textit{ReLU} activations, followed by max-pooling for spatial downsampling. 
We perform global average pooling to reduce the size of the images.
The residual from the previous convolution block is added to the output from the current convolution layer and this is the input to the next convolution block. 
The architecture further includes a convolutional layer and global average pooling to output a one-dimensional tensor. 
The $w$ values are embedded with a dense layer and concatenated with the image representations for joint analysis. 
Regularization techniques, such as dropout and L2 regularization, are applied to the dense layers to prevent overfitting. 
The architecture outputs \textit{logits} which is $\log\left(\frac{p(x|w)}{p(x)}\right)$. 
We use the output from the trained model to estimate the posterior $p(w|x)$.

\begin{figure*}[!ht]
 \centering
    \includegraphics[width=0.6\linewidth]{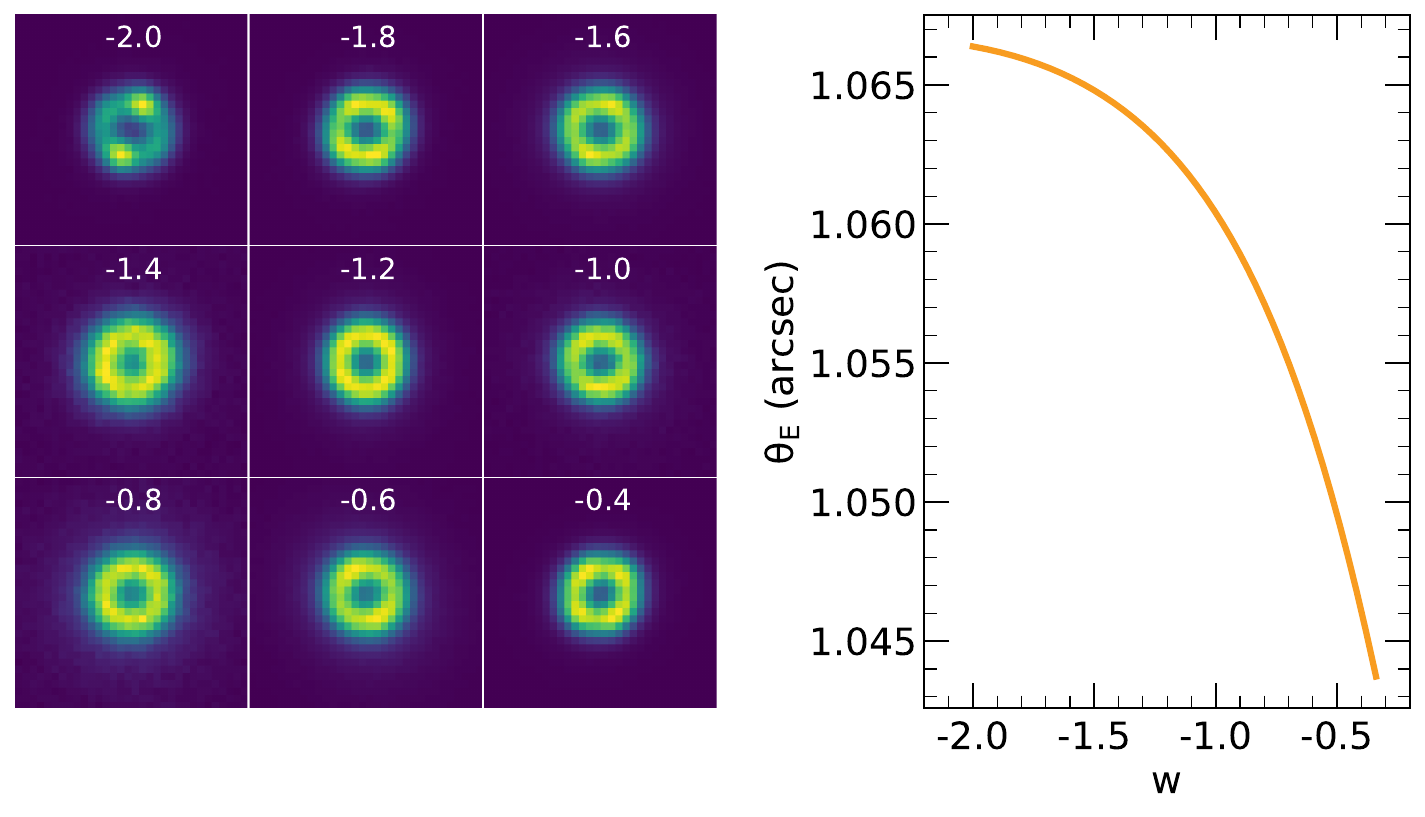}
    \caption{
    Left: Sample of training data with the $w$ value used to generate the image. 
    Right: the Einstein radius $\theta_{E}$ as a function of the dark energy equation-of-state parameter $w$. The variation in $\theta_{E}$ is larger at higher values of $w$ than at the lower values.
    }
    \label{fig:train_data}
\end{figure*}

We train the network by minimizing the BCE loss function by dynamically adjusting the learning rate when the validation loss is plateaued by a decay factor of 0.1 starting from $1e^{-2}$ to $1e^{-6}$, if the validation loss does not improve over five epochs. 
The model is trained over 100 epochs with a batch size of 1024 using \texttt{Adam} optimizer. 
We include an option for early stopping if the validation loss does not improve over 20 epochs. 
The model is trained on NVIDIA A100  GPU  for 71 epochs with a typical training time of 25 minutes.

\subsection{Population-level Posterior Inference} 
\label{sec:posterior}
Given the likelihood-to-evidence ratio, we calculate the posterior using two methods. 
The first is sampling from the joint likelihood ratio using MCMC, and the second is calculating the posterior analytically for each $w$ value drawn from the prior without the need for sampling. 
Using both MCMC and analytical calculation of the posterior from the likelihood-to-evidence ratio serves as complementary methods to validate the accuracy and consistency of the results. 
%We calculate the posterior distributions using these two methods for 5, 100, 500, 100, 2000, and 3000 images.

\subsubsection{Posterior from MCMC} 
\label{sec:mcmc_posterior}
We infer the posterior distributions of $w$ by carrying out MCMC sampling using the Metropolis-Hastings algorithm~\cite{hastings70} from the log likelihood-to-evidence ratio obtained from NRE. We implemented MCMC using the \texttt{emcee} package~\cite{foreman13}. 
The prior range to draw $w$ values is the same as the training prior. 
We initiate five walkers randomly around a starting value of $w=-1.0$. 
We implement a series of two warm-up periods, with 100 and 500 steps respectively, where the walkers converge to the target distribution. 
The position of the walkers is reset to the value that was converged on after each warm-up. 
After the two warm-ups, we run the sampling for 1000 steps. 
The log likelihood-to-evidence ratio $\mathrm{log}\  r(\{x\} | w)$ for a population of images from Equation \ref{eqn:joint_likelihood} is estimated for each $w$ value drawn from the prior using the trained NRE model followed by summation of the ratios. 
%For 3000 images, the MCMC inference takes $\sim$ 30 minutes on A100 NVIDIA GPU.

\subsubsection{Analytical Posterior}
\label{sec:analytical_posterior}

The population-level posterior can be analytically calculated from the individual likelihoods when there are few enough latent parameters --- i.e., when the evidence contains a tractable integral.
% if the parameters of interest is in low dimension which makes it possible to calculate the evidence term in Bayes theorem.

From Equations \ref{eqn:bayes} and \ref{eqn:joint_likelihood}, the posterior $p(w | \{x\})$ from a population of observations $\{x\}$ is given by
\begin{equation}
\begin{split}
    p(w|\{x\}) &= \frac{p(w)~\prod_{i}r(x_i|w)}{\int dw^{\prime}~ p(w^{\prime})~\prod_{i}r(x_{i}|w^{\prime})},\\
        &= p(w)~\left( \int dw^{\prime}~p(w^{\prime})~\prod_{i}\frac{r(x_i|w^{\prime})}{r(x_i|w)} \right)^{-1}.
\end{split}
\end{equation}
We generate a list of $w$ values in the prior range and analytically calculate the above equation for a population of strong lens images at each $w$.

\begin{table*}
\noindent\begin{minipage}[b]{0.99\linewidth}
  \caption{
  Parameter distributions used to generate training and test sets. 
  For the population-level inference, we generate datasets with a fixed $w$ value that is common across all the images.
  Uniform distributions are denoted by $\mathcal{U}$.}
  \label{table:params}
  \centering
  % \begin{tabular}{llrlllll}
  \begin{tabular}{llll}
 \hline   Parameter & Name & Training Priors & Test Priors and Fixed values\\ \toprule  \hline
 \multicolumn{3}{c}{Cosmology}\\
 \midrule
 $w$ & Dark Energy Equation of State &  $\mathcal{U}(-2.0, -0.34)$ & $\mathcal{U}(-2.0, -0.34)$, $-1.2$, $-1.0$, $-0.8$ \\
 $H_{0}$ [km/s/Mpc] & Hubble constant &  70.0 & 70.0 \\
 $\Omega_{de}$ & Dark Matter density & 0.7 & 0.7 \\
 $\Omega_{m}$ & Dark Matter density & 0.3 & 0.3 \\
  \midrule
 \multicolumn{3}{c}{Lens Parameters}\\
 \midrule
  $l_{e_{1}}$  &  lens ellipticity   & $\mathcal{U}(-0.1,0.1)$   & $\mathcal{U}(-0.1,0.1)$  \\ 
  $l_{e_{2}}$  &  lens ellipticity  & $\mathcal{U}(-0.1,0.1)$     &$\mathcal{U}(-0.1,0.1)$ \\ 
 \midrule
 \multicolumn{3}{c}{Source Light Parameters}\\
 \midrule
  $m_s$  & magnitude & $\mathcal{U}(19, 24)$   & $\mathcal{U}(19,24)$ \\ 
  $R$ [arcsec] & half-light radius & $\mathcal{U}(0.1,3.0)$ & $\mathcal{U}(0.1,3.0)$ \\ 
   $n$   & sersic index  & $\mathcal{U}(0.5,8.0)$ & $\mathcal{U}(0.5,8.0)$  \\ 
 $s_{e_1}$  & source ellipticity  & $\mathcal{U}(-0.1,0.1)$  & $\mathcal{U}(-0.1,0.1)$  \\ 
 $s_{e_2}$  & source ellipticity  & $\mathcal{U}(-0.1,0.1)$  & $\mathcal{U}(-0.1,0.1)$ \\ \hline
\end{tabular}
\end{minipage}
\end{table*}

\section{Data}\label{sec:data}
We generate images of strong lenses using \texttt{deeplenstronomy} \citep{morgan21} which is built on \texttt{lenstronomy} \citep{birre15, birrer18}. 
The main components of simulating galaxy-galaxy strong lenses are the lens's mass profile, the source's light profile, and the angular distance diameters between the observer, source, and lens.

We assume that the astrophysics parameters are independent of the cosmology and also independent of each other for simplicity of modeling. 
We also assume that the lens light is perfectly subtracted from the images. 
The lens light is often subtracted before lens modeling and there has been work done on automating this process \cite{hezaveh17}. 
We set the lens and source positions to $(0,0)$,
lens redshift $z_{l} = 0.1$, source redshift $z_{s} = 2.0$, stellar velocity dispersion = 200 $km/s$, $H_{0} = 70\ (km/s)/Mpc$ and $\Omega_{m} = 0.3$. 

In our model, we have:
% \vspace*{-2mm}
\begin{itemize}
    \item observable $x$: strong lens image;
    \item parameter of interest $w$: dark energy equation-of-state parameter; 
    \item nuisance parameters $\nu$: lens ellipticity ($l_{e1}$, $l_{e2}$), magnitude of the source $m_{s}$, source effective radius $R$, source Sersic index $n$, and source ellipticity ($s_{e1}$, $s_{e2}$). While these nuisance parameters are incorporated into the simulation process to generate the images, our objective does not involve inferring their values.
    %These are the parameters which go into the simulation but we do not infer.
\end{itemize}

The strong lens images are generated by sampling from the uniform prior distribution of these parameters as shown in Table \ref{table:params}. 
To generate images representative of the survey conditions, we use DES Data Release 1 \cite{abbott18} with a pixel scale of 0.263 arcsec/pixel, CCD detector gain of 6.083  $e^{-}$/count, and read noise of 7.0 $e^{-}$.  
We simulate $g$-band images with a magnitude zero point of 30.0. 
We include the sky brightness, seeing, and the effective number of exposures based on the survey's empirical values. 
The images have a size of 32 x 32 pixels. 
The training and validation dataset includes 640,000 images and 160,000 images, respectively. 
To assess the model's performance, we use a test dataset of 2,000 images generated from the same prior range as the training dataset. 
For the population-level inference, we create three distinct datasets, each consisting of 3,000 images with a common $w$ value across all images within that dataset. 
Specifically, these three datasets are generated with $w_{\rm true}$ fixed to $-1.2, -1.0$, and $-0.8$, respectively. 
The purpose of these datasets is to assess the capacity of the model to constrain the common value of $w$ simultaneously from multiple observations within a population.
A sample of the training data is shown in Figure \ref{fig:train_data} (left). 
In Figure \ref{fig:train_data}, we show the correlation between the Einstein radius $\theta_E$ and $w$ in the training data: the variation in $\theta_E$ is greater at high $w$ values compared to low $w$. The code and the dataset used in this paper can be found in our github repository \footnote{https://github.com/deepskies/DeepSLIDE}.

\section{Results} 
\label{sec:results}
With the trained NRE model, we infer the cosmological parameter $w$ on a dataset of 2,000 individual strong lenses generated from the same distribution as the training data. 
We also perform population-level inference on a dataset of 3,000 strong lens images that share a common $w$ value. 
This analysis is done to assess the improvement in constraining $w$ by combining information from multiple observations within the population.

\begin{figure*}[!htp!]
    \centering
    \includegraphics[width=0.3\linewidth]
    {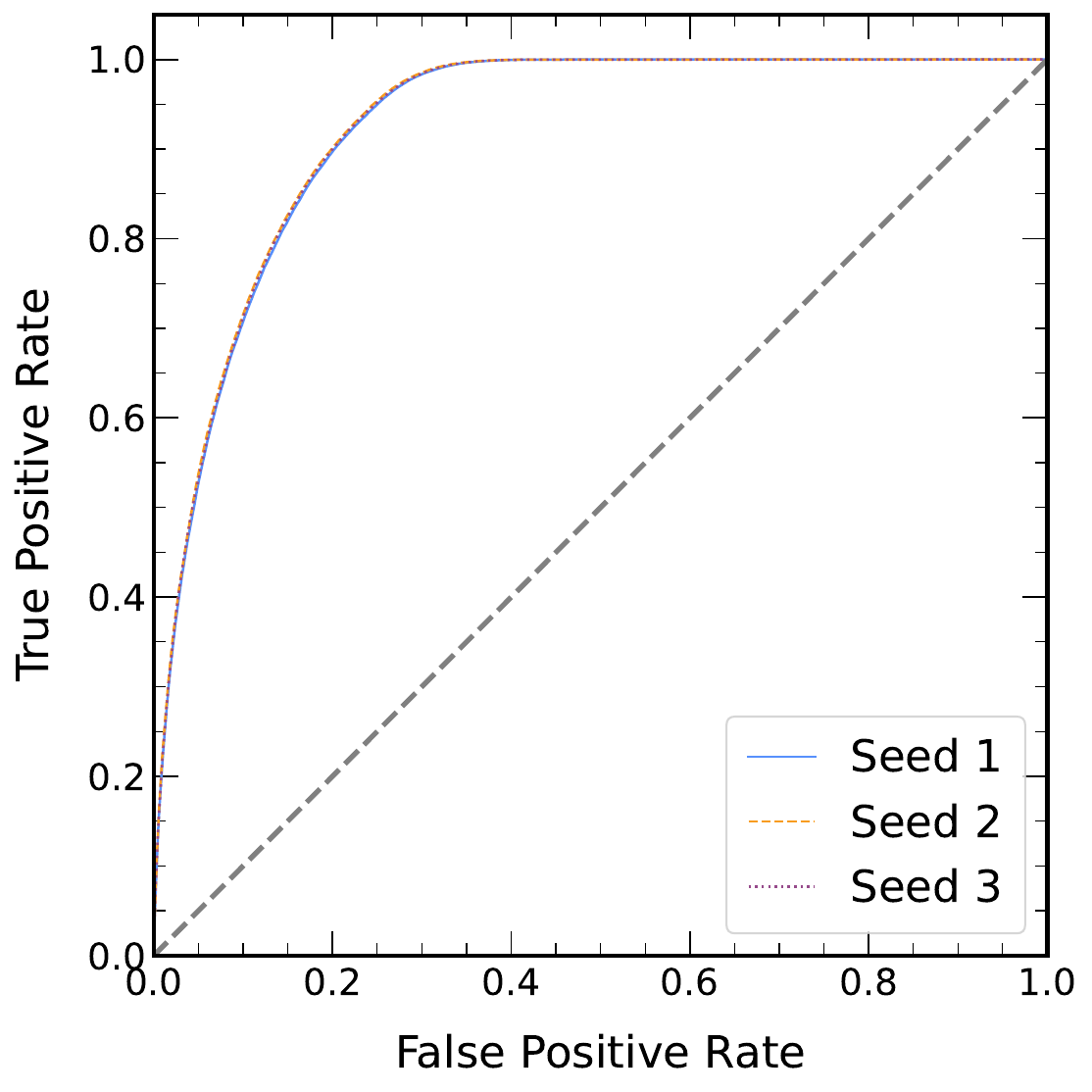}
    \includegraphics[width=0.263\linewidth]{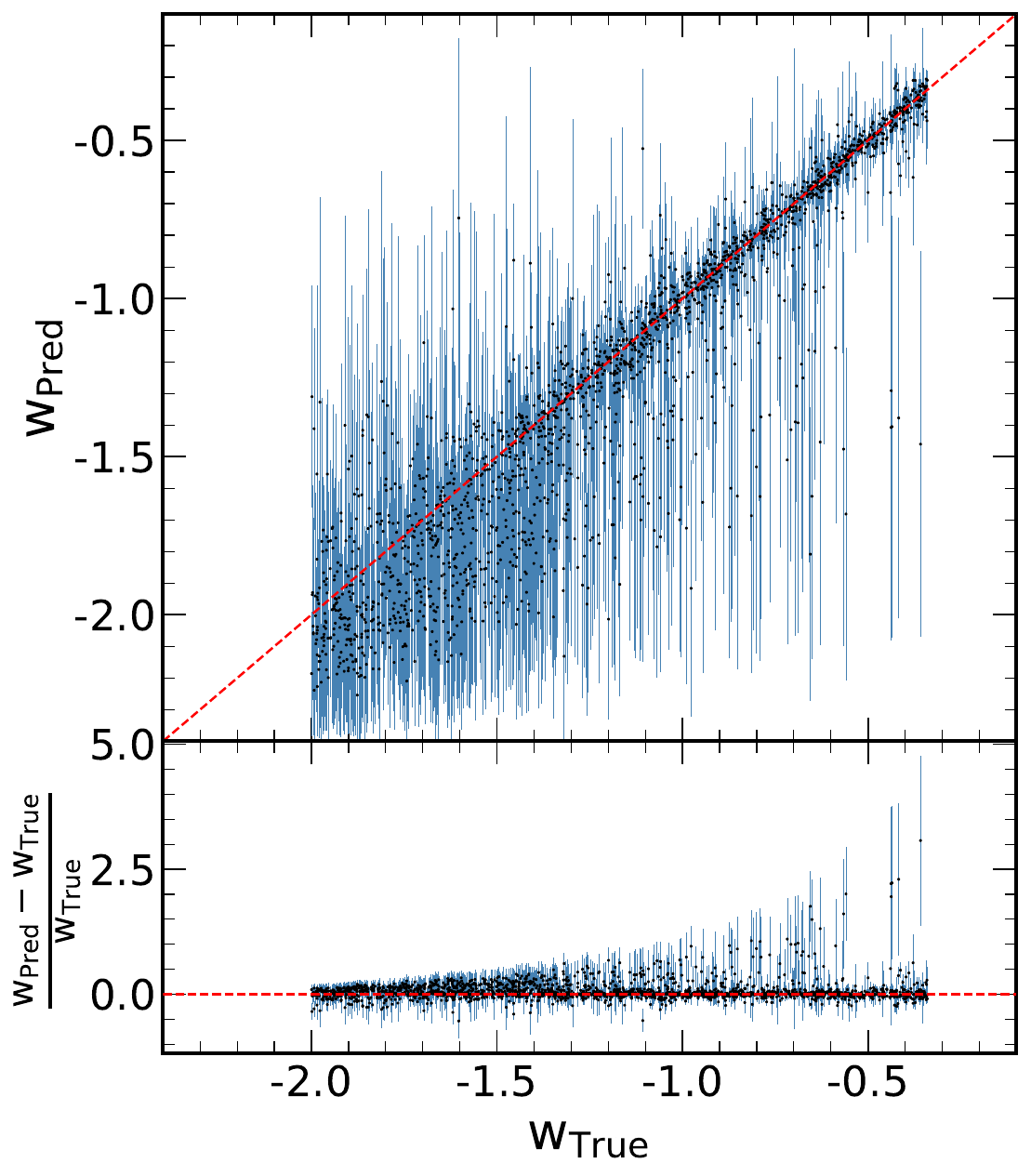}
    \includegraphics[width=0.3\linewidth]{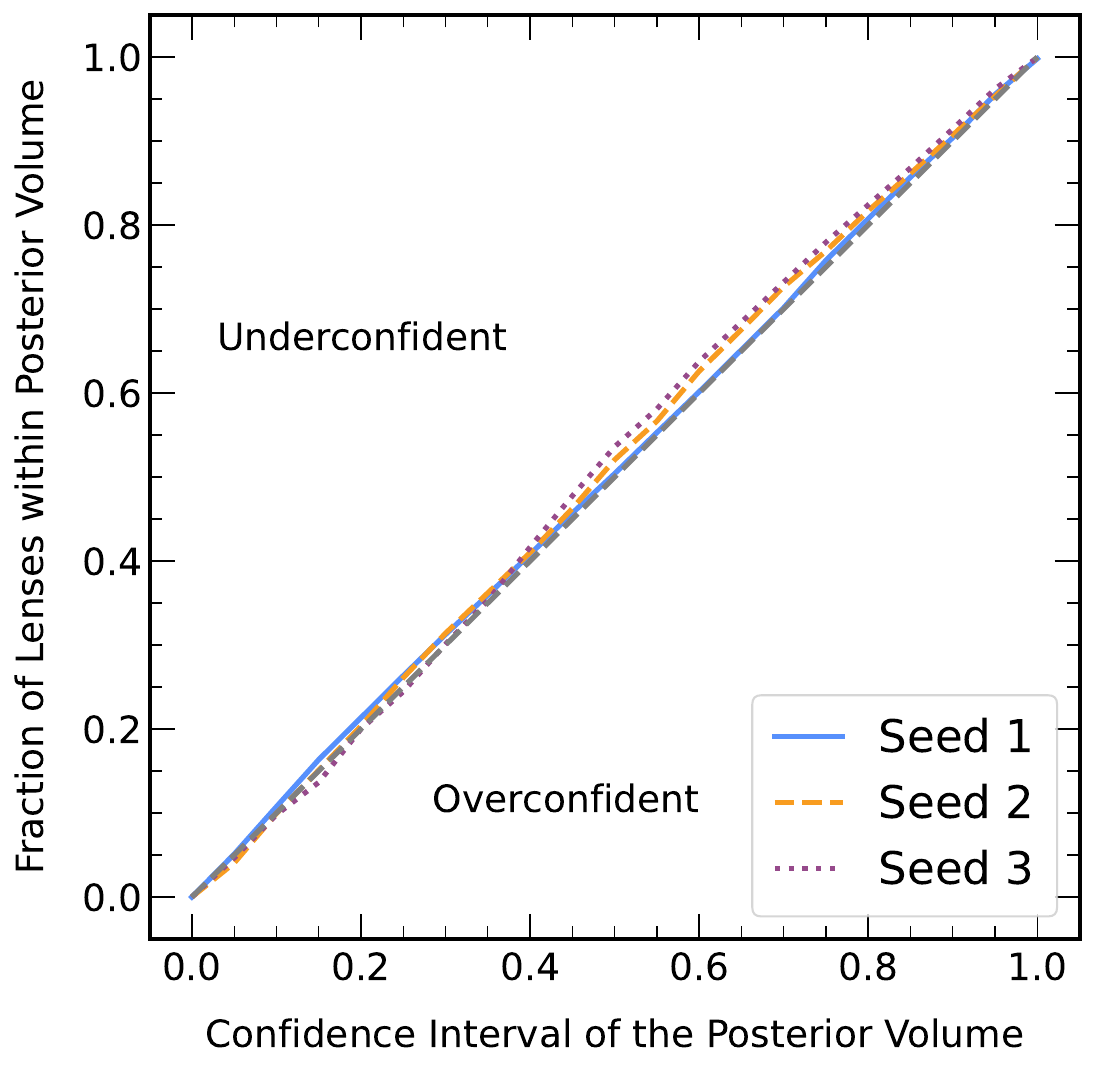}
    \caption{Left: The model performance on 2,000 test images is shown by the Receiver Operating Characteristic (ROC) curve for three models with different seed initialization.
    %The Area under the curve (AUC) is $\sim 0.92$ indicating good model performance across different seed initializations.
    % with a high true positive rate and a low false positive rate. 
    % The ROC curve is similar for all the models indicating the robustness of the models.
    Middle Top: The parity plot showing true Vs predicted $w$ values with the mean and 1$\sigma$ error obtained from the analytical posterior calculation (Section \ref{sec:analytical_posterior}). 
    Middle Bottom: The bias plot with the scaled error. The expected value is shown by the red dashed line. 
    Right: The posterior coverage plot for different weight initialization demonstrating that the model is well-calibrated.}
    \label{fig:roc_auc_parity}
\end{figure*}

We assess the performance of the trained NRE model using the Receiver Operating Characteristic (ROC) curve for different thresholds as shown in Figure \ref{fig:roc_auc_parity} (left). 
The area under the ROC curve (AUC) is $\sim\ 0.92$, which indicates that the model has high discriminating power.
We also check the model's stability by training it on the same training dataset but with three different weight initializations---set by three different random seeds. 
The models from the three seeds have almost identical ROC curves indicating that the model training converges to similar optimal classifiers. 

We calculate the posteriors for the 2,000 images using the analytical method described in Section \ref{sec:analytical_posterior}. 
The analytical posterior approach is computationally less expensive for individual images, especially for large datasets, compared to the MCMC sampling method outlined in Section \ref{sec:mcmc_posterior}.
We present the mean equation-of-state parameter $w_{\rm pred}$ and associated $1\sigma$ uncertainties obtained from the posterior of each image in Figure \ref{fig:roc_auc_parity} (middle, top). 
The figure indicates that $w_{\rm true}$ most often lies within the predicted $1\sigma$ uncertainty for most data points.
However, for observations with $w_{\rm pred} \lesssim -1.4$, the posteriors are less constrained compared to those with $w_{\rm pred} \gtrsim -1.4$. 
We suggest that this is due to the smaller variation in $\theta_{E}$ at low $w_{\rm true}$ than at high $w_{\rm true}$ as seen in Figure \ref{fig:train_data}, which limits the model's capacity to predict features as effectively as it does for high $w_{\rm true}$ values. 
Figure \ref{fig:roc_auc_parity} (middle, bottom) illustrates the bias between the true and predicted values relative to the true value, with error bars scaled by the true value accordingly. 
The low $w_{\rm true}$ values exhibit more bias compared to high $w_{\rm true}$ consistent with the explanation from the model performance in the low $w_{\rm true}$ range.
In future studies, we will examine the training set in more detail and attempt to design priors that mitigate biases in predictions.

Figure \ref{fig:roc_auc_parity} (right) presents
the posterior coverage for three different random seeds for the network weight initialization. 
The posterior coverage is a measure of how well-calibrated the predicted uncertainties of the posteriors are. In a well-calibrated model, the fraction of observations within a posterior volume should match the confidence interval of that volume.
For example, a posterior with 68$\%$
($1\sigma$ equivalent) confidence interval should contain the true value
within that confidence interval 68$\%$of the time. 
Figure \ref{fig:roc_auc_parity} (right) shows that the model uncertainty is well-calibrated for all random seeds initialization --- i.e., all curves follow the diagonal dashed line.

\begin{figure*}[!ht]
    \centering
    \includegraphics[trim={0 0.4cm 0 0},clip, width=0.75 \linewidth]
    {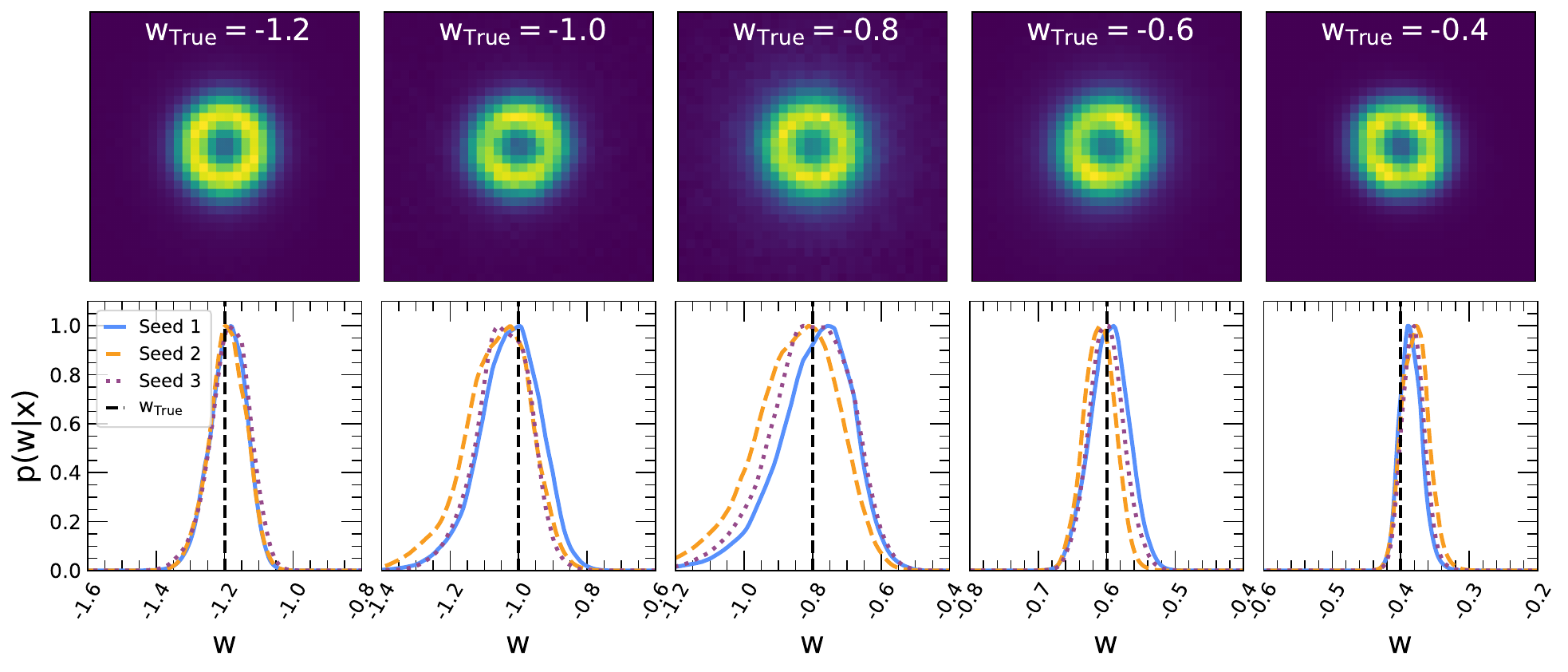}
    \caption{
    Strong lens images (top) and the associated posterior predictions from the analytic method (bottom) with three different random seeds in each case for the neural network weight initialization.
    % Top: Sample images from the test data.
    % Bottom: The analytical posterior $p(w|x)$ calculated from the $\log\left(\frac{p(x|w)}{p(x)}\right)$ from three models with different weight initialization. 
    The black dashed line shows the true $w$ for each inference. 
    %The model is shown to be robust to weight initialization as all three models converge to the same posterior.
    }
    \label{fig:images_posterior}
\end{figure*}

The analytical posteriors for five of the 2,000 images from Figure \ref{fig:roc_auc_parity} are shown in Figure \ref{fig:images_posterior}. 
The posteriors converge to a similar distribution for different random seeds, indicating the model's stability.

% The likelihood ratio predicted by the trained NRE model is used for posterior calculation as described in Section \ref{sec:mcmc_posterior} with MCMC sampling and in Section \ref{sec:analytical_posterior} for analytical calculation. 

% We show that the posterior is better constrained by combining the information from multiple observations than from a single observation.

% \subsection{Inference from an individual strong lens}

\begin{figure*}[!ht]
    \centering
    \includegraphics[width=0.8\linewidth]
    {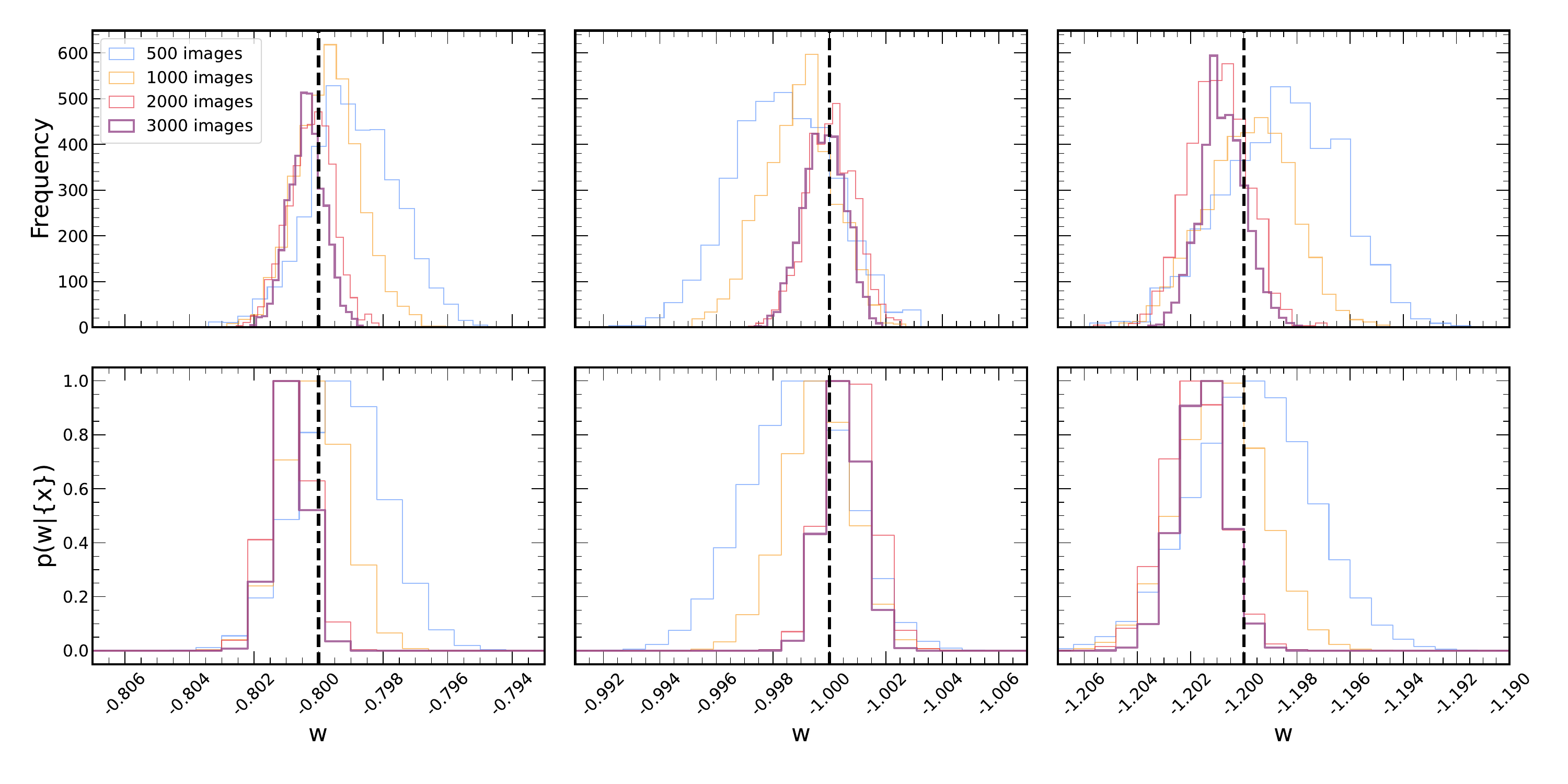}
    \caption{The posterior inference $p(w | \{x\})$ from the joint population analysis of 500, 1000, 2000, and 3000 strong lens images is shown. The posteriors for $w = -0.8, -1.0,$ and $-1.2$ is shown from left to right.
    % The population posterior is calculated by combining the likelihood ratio $p(x_{i}|w)$ for each image $i$. 
    % The true $w$ value is shown in black dashed line.
    Top: The posterior obtained from MCMC sampling (Section \ref{sec:mcmc_posterior}).
    Bottom: The posterior obtained from analytical calculations (Section \ref{sec:analytical_posterior}). In both the methods, the posterior is more constrained as the number of observations in the population increase.
     }
    \label{fig:mcmc_analytical_prob_all}
\end{figure*}

% \subsection{Inference from population of strong lenses}
For the population-level analysis, we estimate the posteriors from the joint likelihood-to-evidence ratio using two complementary methods: MCMC sampling and analytical calculation. 
While MCMC sampling is widely used and particularly efficient in high-dimensional parameter space, it can be computationally expensive. 
In contrast, the analytical calculation of the posterior is a computationally efficient alternative that is suitable in low dimensions. 
Comparing the posteriors obtained from both methods will validate the accuracy and consistency of the population-level inference. 

We perform population-level inference
analysis on three datasets with $w = {-0.8, -1.0, -1.2}$ as described in Section \ref{sec:data}. 
Figure \ref{fig:mcmc_analytical_prob_all} shows the MCMC posterior densities (top panel) and the analytical posterior (bottom panel) for the three $w$ values, each inferred from the joint likelihood-to-evidence ratio of 500, 1000, 2000, and 3000 strong lens observations.
The posterior is better constrained around the true value for the inference performed on the larger population.

\begin{figure}[!ht]
    \centering
    \includegraphics[width=0.8\linewidth]
    {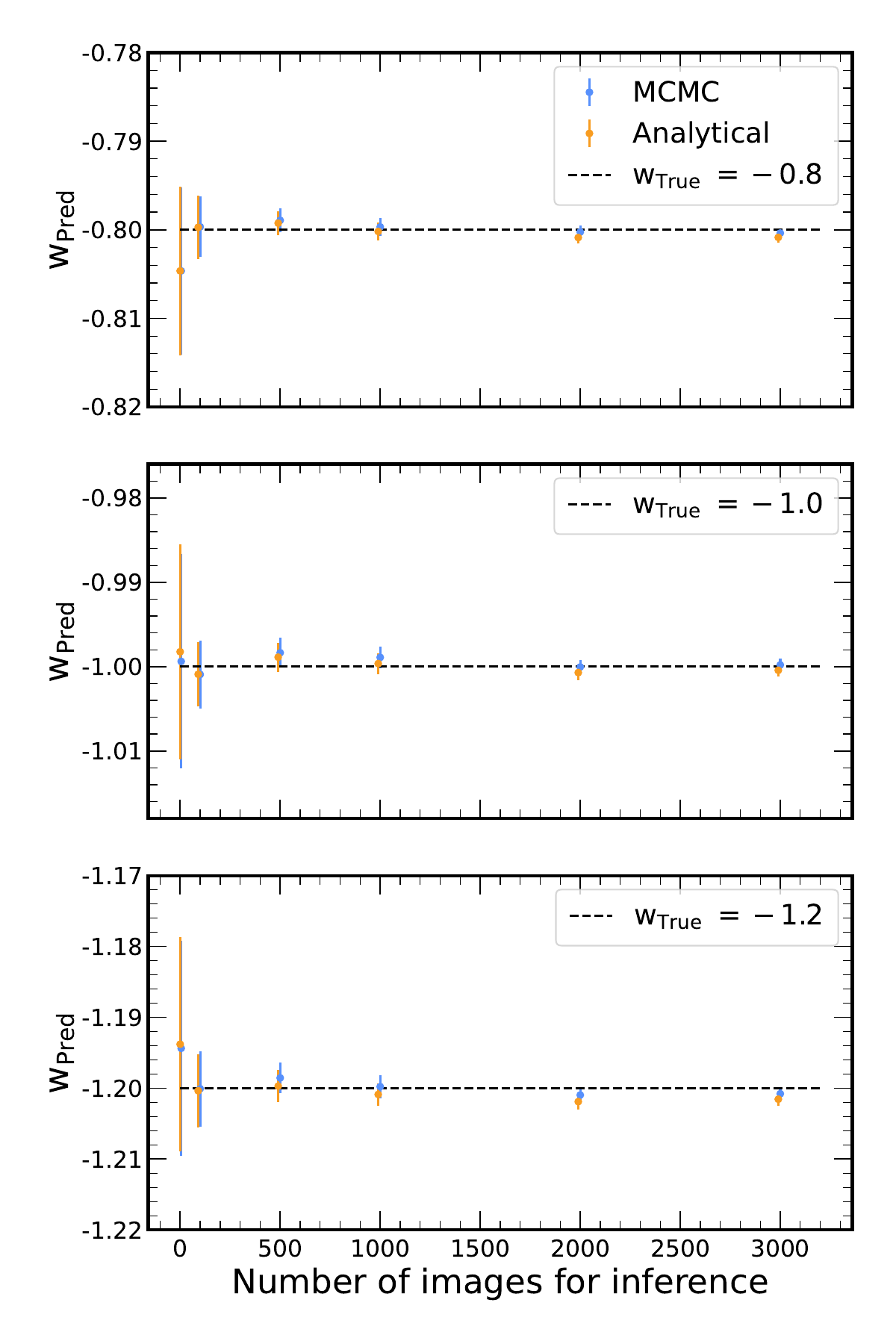}
    \caption{
    Population-level MCMC predictions of the dark energy equation-of-state parameter $w_{\rm pred}$ as a function of the population size (5, 100, 500, 1000, 2000, and 3000 strong lens images) for three different values of $w_{\rm true}$ (-0.8, -1.0, -1.2). 
    % The posteriors obtained from MCMC and from analytical calculation is shown for $w = -0.8, -1.0$, and $-1.2$ from top to bottom respectively. 
    The posterior converges to the true value as the number of observations in the population increases.}
    \label{fig:mcmc_analytical_all}
\end{figure}

Figure \ref{fig:mcmc_analytical_all} presents the mean and $1\sigma$ error of $w$ obtained from MCMC and analytical posterior using the joint likelihood-to-evidence ratio of 5, 100, 500, 1000, 2000, and 3000 images. 
Consistent with the posterior distributions in Figure \ref{fig:mcmc_analytical_prob_all}, the posterior widths of $w$ decrease with increasing population size. 
This behavior is observed for both the MCMC and analytical approaches to estimating the posteriors.
This result highlights the potential gain in precision achievable by combining information from a larger population of observations.

% \begin{figure*}[!ht]
%     \centering
%     \includegraphics[width=0.4\linewidth]
%     {images/MCMC_analytical_posterior_all.pdf}
%     \caption{
%     Population-level MCMC predictions of the dark energy equation-of-state parameter $w_{\rm pred}$ as a function of the population size (5, 100, 500, 1000, 2000, and 3000 strong lens images) for three different values of $w_{\rm true}$ (-0.8, -1.0, -1.2). 
%     % The posteriors obtained from MCMC and from analytical calculation is shown for $w = -0.8, -1.0$, and $-1.2$ from top to bottom respectively. 
%     The posterior converges to the true value as the number of observations in the population increases.}
%     \label{fig:mcmc_analytical_all}
% \end{figure*}

% \begin{figure*}[]
%     \centering
%     \includegraphics[width=0.4\linewidth]
%     {images/MCMC_posterior_w-1.pdf}
%     \includegraphics[width=0.4\linewidth]
%     {images/Analytical_posterior_w-1.pdf}
%     \caption{Left: The posterior distribution from MCMC sampling. Right: The Analytical posterior calculated using equation. True $w = -1.0$.}
%     \label{fig:mcmc_analytical_w-1}
% \end{figure*}

\section{Conclusions and Outlook} 
\label{sec:conclusion}

This study presents the first exploration of the capability of Neural Ratio Estimation (NRE) to constrain the dark energy equation-of-state parameter $w$ from a population of strong gravitational lenses. 

% We confine our study to galaxy-galaxy lenses with relatively simple images --- i.e., they are very close to Einstein Rings (Figure~\ref{fig:train_data}). 
% We simulated strong lensing systems as if observed by the Dark Energy Survey: they include the appropriate noise and image quality settings as described in Section~\ref{sec:data}.
% % The simulations vary several parameters to achieve a large variety of objects. 
% % We fix the redshifts of the lens and source, and the velocity dispersion of the lens. 
% With these simulations, we study the capacity of NRE for posterior inference of $w$ with single lenses and with populations of lenses. 

% For posterior inference via NRE, one first designs and trains a classifier between ... and ..... to obtain neural network-based estimator of Likelihood ratios (LRs).
% Then, one must perform another operation to obtain the posterior $p(w)$ from the LR. 
% We use two approaches to obtain $p(w)$: Analytic and MCMC -- advantages, disadvantages

We use NRE to train a neural network on simulated strong lens images and the corresponding $w$ to estimate the likelihood-to-evidence ratio $\frac{p(x|w)}{p(x)}$. 
We then use this ratio to compute the posterior $p(w|x)$ using two methods---MCMC and analytical calculation. 
The main results and conclusions from our analysis are:

% summarizing results
\begin{enumerate}
    \item We find that the likelihood-to-evidence ratio classifier has high discriminatory power with an AUC  $\sim0.92$. 
    We also find that it is stable to changes in the neural network weight initialization.
    \item From the analytical posterior analysis of 2,000 test images, we find that the true value lies within $1\sigma$ of the posterior for most of the observations. However, the model's performance is less efficient in constraining low values of $w$ compared to higher values, as the variation in the observable Einstein radius is less pronounced at low $w$, limiting the model's ability to learn in that regime. We aim to address this in future work by updating the training distribution to account for the difference in model performance.
    \item From the posterior coverage plots, we find that the model is well-calibrated irrespective of the weight initialization.
    \item We perform population-level analysis by estimating the posterior using the joint likelihood-to-evidence ratio. Comparing the joint posterior for $w$ values of -0.8, -1.0, and -1.2 obtained through both MCMC and analytical calculation, we observe that the posterior width decreases with an increasing number of observations in the inference population.
\end{enumerate}

From our analysis, we demonstrate that the dark energy equation-of-state parameter is better constrained from a population of strong lenses than from individual images. 
In this proof-of-concept analysis, we use a simplified dataset designed with sufficient variation for training: fixed lens redshift, source redshift, and stellar velocity dispersion; and varying the lens ellipticity, source magnitude, half-light radius, Sersic index, and source ellipticity. 
% The classifier model has a relatively simple structure with a low number of parameters.
Improvements upon our study could include inference from more complex and realistic images representing different configurations of sources and lenses, testing observational aspects of other surveys, and performing a joint inference of $w$ and $\Omega_{m}$. 
With upcoming surveys anticipated to discover thousands of strong lenses, there is a significant opportunity for scientific discovery. 
To fully utilize the data from these surveys, we must harness scalable population-level inference methods, such as the one demonstrated in this analysis.

\section{Acknowledgements}
This manuscript has been supported by Fermi Research Alliance, LLC under Contract No. DE-AC02-07CH11359 with the U.S. Department of Energy (DOE), Office of Science, Office of High Energy Physics. We acknowledge the Deep Skies Lab, a community of experts and collaborators who played an important for the development of this project. 

% Traditional methods to model each strong lens with a lot of latent variables (or nuisance parameters which are part of the simulation) individually is computationally expensive and challenging. 
% Machine Learning Simulation-Based inference methods such as Neural likelihood ratio estimation (NRE) allows us to estimate the approximate likelihood ratio by marginalizing over the latent variables. 
% This algorithm also alows us to get the population level inference of a common parameter of interest by combining the likelihood ratios from individual observations.

% In this analysis, we use NRE to infer the dark energy equation of state parameter $w$ from a population of strong lens images. 

% We hope to implement our model on the observations from the upcoming strong lens follow-up surveys such as the 4MOST Strong Lensing Spectroscopic Legacy Survey (4SLSLS) which is expected to observe 10000 strong lenses. 
% To implement our model on real data, we aim to train our model on realistic observations with redshifts and velocity drawn from a prior distribution with real noise. 
% We will perform a joint inference of the dark energy equation of state parameter $w$ and the dark matter density parameter $\Omega_{m}$ to compare with the existing literature using MCMC analysis.

\clearpage

%%%%%%%%%%%%%%%%%%%%%%%%%%%%%%%%%%%%%%%%%%%%%%%%%%%%%%%%%%%%%%%%%%%%%
%%%%%%%%%%%%%%%%%%%%%%%%%%%%%%%%%%%%%%%%%%%%%%%%%%%%%%%%%%%%%%%%%%%%%
%\clearpage
\bibliography{main}
\bibliographystyle{icml2024}

%%%%%%%%%%%%%%%%%%%%%%%%%%%%%%%%%%%%%%%%%%%%%%%%%%%%%%%%%%%%%%%%%%%%%
%%%%%%%%%%%%%%%%%%%%%%%%%%%%%%%%%%%%%%%%%%%%%%%%%%%%%%%%%%%%%%%%%%%%%
% APPENDIX
%%%%%%%%%%%%%%%%%%%%%%%%%%%%%%%%%%%%%%%%%%%%%%%%%%%%%%%%%%%%%%%%%%%%%
%%%%%%%%%%%%%%%%%%%%%%%%%%%%%%%%%%%%%%%%%%%%%%%%%%%%%%%%%%%%%%%%%%%%%
\newpage
\appendix
\onecolumn

\section{NRE Network Architecture} \label{sec:app:network}
The architecture for NRE is shown in Table \ref{tab:network_summary_skip}.

\begin{table}[h]
\centering
\begin{tabular}{lll}
\hline
Layer Type & Output Shape & No: of Parameters \\\hline
Input Images $x$ & (Batch, 32, 32, 1) & 0 \\
2D Convolution & (Batch, 16, 16, 8) & 80 \textcolor{teal}{$\leftarrow$}\\
2D Convolution & (Batch, 16, 16, 16) & 1168 \\
2D Convolution & (Batch, 16, 16, 16) & 2320 \\
2D MaxPooling & (Batch, 8, 8, 16) & 0 \\
2D Convolution & (Batch, 8, 8, 16) & 144 \\
Add & (Batch, 8, 8, 16) & 0 \textcolor{teal}{$\leftarrow$} \\
2D Convolution & (Batch, 8, 8, 32) & 4640 \\
2D Convolution & (Batch, 8, 8, 32) & 9248 \\
2D MaxPooling & (Batch, 4, 4, 32) & 0 \\
2D Convolution & (Batch, 4, 4, 32) & 544 \\
Add & (Batch, 4, 4, 32) & 0 \textcolor{teal}{$\leftarrow$} \\
2D Convolution & (Batch, 4, 4, 45) & 13005 \\
2D Convolution & (Batch, 4, 4, 45) & 18270 \\
2D MaxPooling & (Batch, 2, 2, 45) & 0 \\
2D Convolution & (Batch, 2, 2, 45) & 1485 \\
Add & (Batch, 2, 2, 45) & 0 \textcolor{teal}{$\leftarrow$} \\
2D Convolution & (Batch, 2, 2, 64) & 3349 \\
Input $w$ & (Batch, 1) & 0 \\
GlobalAveragePooling & (Batch, 64) & 0 \\
Dense & (Batch, 64) & 128 \\
Dense & (Batch, 128) & 16512 \\
Dropout & (Batch, 128) & 0 \\
Dense & (Batch, 64) & 8256 \\
Dropout & (Batch, 64) & 0 \\
Dense (log $r$) & (Batch, 1) & 65 \\\hline
\end{tabular}
\caption{Summary of the network architecture with residual skip connections indicated by the teal arrows which are added through the \textit{Add} layer. Batch normalization and activation functions are added to each of the convolution and dense layers.}
\label{tab:network_summary_skip}
\end{table}

\end{document}